\documentclass[12pt,preprint,amssym]{aastex}
\slugcomment{Accepted for Publication in the Astrophysical Journal}

\shortauthors{Kulkarni et al.}
\shorttitle{Damped Ly-$\alpha$ Absorbers}
\begin{document}

\title{Metallicity Evolution of Damped Lyman-Alpha Galaxies}

\author{Varsha P. Kulkarni}
\affil{Dept. of Physics \& Astronomy, University of South Carolina,  
Columbia, SC 29208}
\author{S. Michael Fall}
\affil{Space Telescope Science Institute, 3700 San Martin Drive, Baltimore, 
MD 21218}

\vskip 1.0in

\begin{abstract}
We have reanalyzed the existing data on Zinc abundances in damped 
Ly$\alpha$ (DLA) absorbers to investigate whether their mean metallicity 
evolves with time. Most models of cosmic chemical evolution predict that 
the mass-weighted mean interstellar 
metallicity of galaxies should rise with time from a low value $\sim 1/30$ 
solar at $z \sim 3$ to a nearly solar value at $z \sim 0$. However, several 
previous analyses have suggested that there is little or no evolution in the 
global 
metallicity of DLAs. The main problem is that the 
effective number of systems that dominate the $N({\rm H \, I})$-weighted 
mean metallicity is very small. 

We have used a variety of statistical techniques to quantify the global  
metallicity-redshift relation and its uncertainties, taking into account 
both measurement and sampling errors. Three new features of our 
analysis are: (a) an unbinned 
$N({\rm H \, I})$-weighted nonlinear $\chi^{2}$ fit to an exponential relation; 
(b) survival analysis to treat the large number of limits in the existing 
data; and (c) a comparison of the data with several models of 
cosmic chemical evolution based on an unbinned $N({\rm H \, I})$-weighted 
$\chi^{2}$. We find that 
a wider range of evolutionary rates is allowed by the present data than 
claimed in previous studies. The slope of the exponential fit to the 
$N({\rm H \, I})$-weighted  mean Zn metallicity 
vs. redshift relation is $ -0.20 \pm 0.11$ counting limits as 
detections and $ -0.27 \pm 0.12$ counting limits as zeros. 
Similar results are also obtained if the data are binned 
in redshift, and if survival analysis is used. These slopes are marginally 
consistent with no evolution, but are also consistent 
with the rates predicted by several models of cosmic chemical evolution 
(e.g., slopes of $-0.61$ to $-0.25$ for the models of Pei \& Fall 1995, 
Malaney \& Chaboyer 1996, and Pei et al. 1999). The $\chi^{2}$ 
values obtained for most of these models are somewhat worse than that for the  
exponential model because the models lie above the observed data points, 
but still suggest that the present DLA data could indicate some evolution of 
the metallicity with redshift. Finally, we outline 
some future measurements necessary to improve the statistics of the 
global metallicity-redshift relation. 
\end{abstract}

\keywords{quasars: absorption lines; galaxies: evolution; 
galaxies: abundances; cosmology: observations}

%

\newpage
\section{INTRODUCTION} 
The evolution of stars and gas in galaxies are topics of great interest in 
modern astrophysics. The average star-formation history of the 
universe has been estimated from emission properties 
of galaxies detected in deep imaging and redshift surveys 
(e.g. Lilly et al. 1996; Madau et al. 1996, 
1998). This emission history of galaxies is connected 
intimately with the histories of gas consumption and metal production, 
because the global densities of gas, metals, and stars are coupled through 
conservation-type relations (e.g., Fall 2001). In particular, because the 
global rate of star formation is known to be high at 
$1 \lesssim z \lesssim 4$, we expect 
the mean interstellar metallicity of galaxies to rise rapidly in that 
interval. Direct observational constraints on the evolution of the mean 
metallicity 
in galaxies are therefore important for pinning down the histories  
of star formation and gas consumption.

Abundance measurements in gas traced by quasar absorption lines can directly 
probe 
the evolution of metals in galaxies. The damped Ly$\alpha$ (DLA) 
absorbers (log $N({\rm H \, I}) \gtrsim 20$) are particularly 
important, since they contain a large fraction  
of the neutral gas in galaxies, nearly enough to 
form all of the stars visible today (e.g., Wolfe et al. 1995). 
DLAs are the only class of high-redshift objects in which abundances of 
a large number of elements have been measured   
(e.g., Meyer \& York 1992; Pettini et al. 1994, 1997, 
1999, 2000; Lu et al. 1996; Kulkarni et al. 1996, 1997, 1999; Prochaska \& 
Wolfe 1996, 1997, 1999; Prochaska et al. 2001b; and references therein). In 
previous studies, Zn II has been the most commonly used ion 
for estimating the total (gas + solid phase) metallicity 
in DLAs, for a number of reasons: (1) Zn is relatively 
undepleted on interstellar dust grains; (2) it tracks Fe closely 
in Galactic stars (for [Fe/H] $\gtrsim -2$); (3) its absorption lines are 
usually unsaturated and often lie outside the Ly$\alpha$ forest; and 
(4) ionization corrections are relatively small for Zn II. 
Abundances of depleted elements such as Cr or Fe relative to Zn are 
used to estimate the dust content of DLAs. 

The quantity of interest here is the $N({\rm H \, I})$-weighted mean 
metallicity, which corresponds to the global  
$\Omega_{\rm{metals}}^{\rm ISM}/\Omega_{\rm{gas}}^{\rm ISM}$ ratio in 
galaxies. 
The average interstellar properties of galaxies can be determined from 
the statistics of quasar absorption-line systems as follows.
Let $f(N_x,z)$ be the distribution in column density and redshift, 
for particles of any
type $x$ that absorb or scatter light.
These might, for example, be hydrogen atoms ($x=$~H I), metal ions ($x=m$), 
or dust grains ($x=d$). 
By definition, $H_0 (1+z)^3 |dt/dz| f(N_x,z)dN_xdz$ is the mean number 
of absorption-line systems with column densities of $x$ between $N_x$ and 
$N_x+dN_x$ and redshifts between $z$ and $z+dz$ along the lines of sight 
to randomly selected background quasars.
These lines of sight are very narrow (the projected size of the continuum-
emitting regions of quasars, less than a light-year across) and pierce 
the absorption-line systems at random angles and impact parameters.
One can show that the mean comoving density of $x$ is given by 
\begin{equation}
\Omega_x(z) = {8 \pi G m_x \over 3 c H_0} \int_0^\infty N_x f(N_x,z) dN_x,
\end{equation}
where $m_x$ is the mass of a single particle (atom, ion, or grain).
Equation~(1) plays a central role in this subject.
It enables us to estimate the mean comoving densities of many quantities
of interest without knowing anything about the structure of the 
absorption-line systems.
In particular, we do not need to know their sizes or shapes, whether they
are smooth or clumpy, and so forth. Furthermore, in the absence of selection 
biases, equation (1) gives exactly the correct weighting to lines of sight 
at different impact parameters, i.e. distances from the centers of 
galaxies. This point is sometimes confused in 
the literature and even the opposite is sometimes asserted (e.g., Edmunds 
\& Phillips 1997). 
A corollary of equation~(1) is that the global interstellar metallicity, 
$Z\equiv\Omega_{\rm metals}^{\rm ISM}/\Omega_{\rm gas}^{\rm ISM}$, is given 
simply by an average over 
the metallicities of individual absorption-line systems weighted by their 
H I column densities.

Most models of cosmic chemical evolution predict that the global interstellar 
metallicity rises with time, from a low value at high redshifts to a 
near-solar value at zero redshift (e.g., Lanzetta, Wolfe, \& Turnshek 1995; 
Pei \& Fall 1995; Malaney \& Chaboyer 1996; Pei, Fall, \& Hauser 1999; 
Tissera et al. 2001). Moreover, from emission-line observations 
of nearby galaxies, it 
appears that the present-day mass-weighted mean interstellar metallicity 
of galaxies, averaged over all morphological types, is near-solar.  
We show this explicitly in the Appendix using the mean relation between galaxy 
luminosity and interstellar metallicity as derived from H II regions, and 
integrating over the luminosity function of galaxies. 
However, it is not clear from the absorption-line observations  
whether or not the mean  
metallicity in DLAs actually increases with decreasing redshift as 
predicted by the models.

There have been several attempts to estimate the $N({\rm H \, I})$-weighted  
mean Zn metallicity of DLAs  
at $0.4 \lesssim z \lesssim 3.5$ (e.g., Pettini et al. 1997, 1999; 
Vladilo et al. 2000; Savaglio 2001). These studies have claimed that there 
is little or no evolution in the global 
Zn metallicity of DLAs in this redshift range. Prochaska \& Wolfe 
(1999, 2000, 2002) and Prochaska, Gawiser, \& Wolfe (2001a) have found no 
evolution in the mean Fe abundance for $2 \lesssim z \lesssim 4$. 
However, these previous studies have not made consistent quantitative 
estimates of the mean slope of the global metallicity-redshift relation 
and the observational uncertainties in that relation. There are, in fact,  
several uncertainties in the present DLA metallicity data. First of 
all, the metallicities for individual 
DLAs show a large intrinsic or cosmic scatter at any given redshift, 
reflecting the different rates of chemical enrichment of different galaxies. 
Furthermore, the current samples are relatively 
small. More importantly, the estimates of the $N({\rm H \, I})$-weighted 
mean metallicity for these samples are dominated by only the 
few DLAs with the highest $N({\rm H \, I})$. To discriminate between evolution 
vs. no evolution from these intrinsically noisy 
samples, it is crucial to estimate quantitatively the mean 
metallicity-redshift relation and the effect 
of the observational uncertainties on that relation. Here we present 
a reexamination of the Zn data with quantitative estimates  
of the mean metallicity-redshift relation and comparison with predictions 
of cosmic chemical evolution models. We include several 
recent low-redshift measurements, which are crucial for determining the 
slope of the metallicity-redshift relation. Using a number of different 
statistical techniques, we find that the existing data could 
support significant evolution of the global metallicity with redshift. 

\section{ANALYSIS AND RESULTS}

To set the stage for our analysis, we begin by giving an overview of the 
previous studies of the global metallicity-redshift 
relation for the DLAs. Table 1 compares the different studies in terms of 
the elements used, the number of detections and upper limits included, 
the way the limits were treated, the redshift range covered, whether 
the data were binned, whether the 
errors in the $N({\rm H \, I})$-weighted means were calculated consistently, 
whether the slope was quantified, and 
finally, the conclusions that were reached regarding metallicity 
evolution. Here we have not included studies of unweighted mean 
metallicities, since these are not appropriate for estimates 
of the global metallicity-redshift relation. Savaglio (2001) has 
reported that the subsample of low-$N({\rm H \, I})$ DLAs shows some 
metallicity evolution. We do not include this result in Table 1 since 
such a subsample does not give the total global metallicity. 

It is clear from Table 1 that 
most previous studies did not determine the slope of the global 
metallicity vs. 
redshift relation. The only one that did estimate the slope 
underestimated the uncertainties (by giving equal weight to each DLA 
while calculating the errors in the $N({\rm H \, I})$-weighted 
mean metallicities). Our study (summarized in the last line of 
Table 1) presents a more rigorous and quantitative analysis of the data and 
shows that contrary to the no-evolution picture suggested by most researchers 
in this field, the current data are actually consistent with   
substantial evolution of the metallicity with redshift.  

For this study, we use only Zn measurements, because this provides us 
with a large sample spanning a relatively wide redshift range 
and allows us to analyze the data uniformly without having to correct 
for dust depletion. (See section 3 for a more detailed 
discussion of this point.) Using Zn alone and excluding Fe does 
limit the size 
of our samples at $z > 3$, but that redshift range corresponds to only 
$16 \, \%$ of the age of the universe \footnote{Throughout this article, 
we compute fractions of the age of the universe using the ``concordance'' 
cosmological parameters $\Omega_{m} = 0.3$ 
and $\Omega_{\Lambda} = 0.7$.}. We do not include DLAs 
with $z_{\rm abs} \approx z_{\rm em}$ in broad absorption line (BAL) quasars, 
since these are thought to contain matter outflowing from the quasars, 
rather than typical galactic interstellar matter. We also eliminate  
DLAs with uncertain or unavailable H I column densities. 
Our samples include Zn measurements from Meyer \& York (1992); Meyer et al. 
(1995); Lu et al. (1995, 1996); Boisse et al. (1998); Pettini et al. 
(1994, 1997, 1999, 2000); Prochaska \& Wolfe (1996, 1997, 
1999); Lopez et al. (1999, 2002); de la Varga et al. (2000); 
Petitjean, Srianand, \& Ledoux (2000); Molaro et al. (2001); 
Ellison \& Lopez (2001); Prochaska et al. (2001b); 
and Levshakov et al. (2002). 
In addition, we also include results from 
a survey of DLA abundances for $0.6 < z < 2.3$, performed with 
the Multiple Mirror 
Telescope (Kulkarni, Bechtold, \& Ge 1999; Ge, Bechtold, \& Kulkarni 2001). 
This is the complete list of published measurements, to our 
knowledge, as of 2002 June 20. Together 
these represent 36 detections and 21 limits. More than $80 \%$ of the 
limits are 3 $\sigma$, while the rest are 4 $\sigma$. 
The interested reader can access our samples at 
http://boson.physics.sc.edu/$\sim$kulkarni/DLAZndata.html. 

We now describe our analysis 
for the trends in the global metallicity based 
on these samples. We assume throughout that Zn is not depleted and is a good 
indicator of 
metallicity. Furthermore, we assume that most of the hydrogen in the 
DLAs is neutral and atomic (H I) while most of the Zn is singly ionized 
(Zn II). These are standard assumptions also made in previous studies in 
this field. The metallicity of an individual DLA is thus approximated by  
\begin{equation}
Z_{i} =  {N({\rm Zn \, II})_{i} / N({\rm H \, I})_{i} \over 
{({\rm Zn/H})_{\odot}}} Z_{\odot}, 
\end{equation}
where $Z_{\odot} = 0.02$ is the present-day solar metallicity, and 
$({\rm Zn/H})_{\odot} 
= 10 ^{-7.35}$ is the solar Zn fraction (Anders \& Grevesse 1989). 
The $N({\rm H \, I})$-weighted mean metallicity $\overline{Z}$ in a sample 
of DLAs is then given by  
\begin{equation}
\overline{Z} = 
{\Sigma N({\rm Zn \, II})_{i} / \Sigma N({\rm H \, I})_{i} \over 
{({\rm Zn/H})_{\odot}}} Z_{\odot}.   
\end{equation}
This can be reexpressed in the more familiar form of a weighted mean 
\begin{equation}
\overline{Z} = \Sigma w_{i} Z_{i}, 
\end{equation}
where the normalized weighting factors 
\begin{equation}
w_{i} = N ({\rm H \, I})_{i}/ \Sigma N ({\rm H \, I})_{i}
\end{equation}
are the fractional contributions of the individual DLAs to the total 
H I column density of the sample. We quantify the evolution of 
$\overline{Z}$, first 
without binning the data, and then with a binned approach for the purpose 
of pictorial presentation. 

\subsection{The Unbinned Approach}

Here, we fit a simple model $\overline Z_{\rm p} (z)$ to the individual 
metallicities $Z_{i}$ by the method of weighted nonlinear least squares. 
This is the first study to use such an unbinned approach for the analysis of 
DLA metallicity evolution. Thus, we minimize the quantity  
\begin{equation}
\chi^{2} = \Sigma w_{i} \biggl [{ {Z_{i} - \overline Z_{\rm p} (z_{i})} \over 
{\sigma_{Z}}} \biggr ] ^{2}, 
\end{equation}
where the weights $w_{i}$ are given by equation 
(5). It can be 
verified that minimization of the $\chi^{2}$ thus defined 
gives the same results as a binned approach in the hypothetical limiting case 
of a flat mean metallicity-redshift relation,  
$\overline Z_{\rm p} (z) = $ const.

For the sake of definiteness, we assume that the models 
$\overline Z_{\rm p} (z)$ 
can be specified with two parameters, and denote by 
$\overline Z_{\rm p}^{\rm {best}} (z)$ the best-fit model 
corresponding to those values of the 
parameters that minimize the $\chi^{2}$. 
The quantity $\sigma_{Z}$ in equation (6) is a measure of the 
scatter among the observed metallicities $Z_{i}$ with respect to the mean 
trend. {\footnote{Note that we cannot use the measurement errors to calculate 
$\sigma_{Z}$, because they are small compared to the scatter in the data.}}  
We take $\sigma_{Z}$ as 
the scatter with respect to the best-fit model:  
\begin{equation}
\sigma_{Z}^{2} = 
{{\Sigma w_{i} [Z_{i} - \overline Z_{\rm p}^{\rm {best}} (z_{i})]^{2}} 
\over {(n-2)}}.
\end{equation}
Thus, by definition, the best-fit model has a reduced chi-square  
($\chi^{2}_{\nu}$) of 1. It is, therefore, not possible to get an independent 
estimate of the goodness of fit. But the contour corresponding to 
$\chi^{2} = \chi^{2}_{\rm min} + 1$ can be used to obtain error bars on 
the parameters of the model $\overline Z_{\rm p} (z)$. 

\subsubsection{Exponential Models}

The few previous studies that have attempted to quantify the observed 
shape of 
the metallicity-redshift relation have fitted a straight line to 
the logarithm of the mean metallicity vs. redshift data, of the form 
\begin{equation}
{\rm log} \,\overline Z = {\rm log} \,\overline Z_{0} + bz.
\end{equation}
The corresponding expression for the mean metallicity itself has an 
exponential dependence on redshift: 
\begin{equation} 
\overline Z_{\rm p} (z) =  \overline Z_{0} \, 
{\rm exp} (-a z),  
\end{equation}
where $a = -b \, {\rm ln}(10)$.  
In fact, as we will discuss below and illustrate in Fig. 1, 
most models of cosmic chemical evolution also predict a roughly exponential 
increase in the mean metallicity with 
decreasing redshift over the range of the DLA Zn observations. 
Therefore, we begin our analysis by assuming an exponential form 
for the predicted metallicity for easy comparison with previous studies. 
We then follow this in section 2.3 with more detailed comparisons with the 
predictions of several models of cosmic chemical evolution. 
 The best-fitting values of the parameters ${\overline Z_{0}}$ and 
$a$ can be obtained by minimizing $\chi^{2}$  numerically, using 
equations (6) and (9). 
The 1 $\sigma$ errors in ${\overline Z_{0}}$ and $a$ can then 
be estimated using equations (6), (7), and (9), from 
the contour in the ${\overline Z_{0}}-a$ plane corresponding to 
$\chi^{2} = \chi^{2}_{\rm min} + 1$. 

The situation is complicated somewhat by the large number of upper 
limits on the Zn~II column densities in the current samples. 
Although a rigorous way to perform linear unweighted regression with 
both detections and limits is using survival analysis, to our knowledge, 
no statistical techniques have been developed as yet for nonlinear 
weighted regression using survival analysis. Therefore, we consider 
two extreme cases to take account of the upper limits. We first treat 
the upper limits as detections, and 
obtain the intercept ${\rm log} \,(\overline{Z}_{0}/Z_{\odot}) = 
-0.71 \pm 0.20$ 
and the slope 
$b = -0.20 \pm 0.11$ for the logarithmic mean metallicity vs. linear 
redshift relation. In the opposite extreme, we treat the upper 
limits as zeros and obtain  
$ {\rm log} \,(\overline{Z}_{0}/Z_{\odot}) = -0.64 \pm 0.20$ and 
$b = -0.27 \pm 0.12$. (The errors in the slope and intercept are, 
of course, closely correlated.) These results are summarized in Table 3. 
The slopes differ from zero at $\approx 2 \sigma$ levels, suggesting 
that the present data could support evolution of 
the mean metallicity with redshift. The intercepts 
${\rm log} \,(\overline{Z}_{0}/Z_{\odot})$ differ from zero at $> 3 \, \sigma$ 
levels. But this is not a serious problem because the intercepts 
involve an extrapolation outside the redshift range of the available data.

\subsection{The Binned Approach}

We now consider the binned approach for the purpose of pictorial 
presentation. We divide the sample into five redshift bins with roughly equal 
numbers of DLAs per bin. 
For each redshift bin, the error in the $N({\rm H \, I})$-weighted mean 
metallicity $\overline{Z}$ can be calculated by using 
the standard expressions for the weighted mean and its sampling 
and measurement errors 
(e.g., Bevington \& Robinson 1992). The Poisson statistical 
error in the weighted mean, because of the finite size of the sample, 
is given by  
\begin{equation}
\sigma^{2}(\overline{Z})_{\rm samp} = 
{{\Sigma [w_{i} (Z_{i} - \overline{Z})^{2}]} 
\over {(n-1) }},  
\end{equation}
where $n$ is the number of measurements in a given bin. 

It is informative to note that equation (10) can also be rewritten as 
\begin{equation}
\sigma^{2}(\overline{Z})_{\rm samp} =
{\Sigma (Z_{i} - \overline{Z}^{'})^{2} \over {n_{\rm eff} (n_{\rm eff} -1)}},
\end{equation}
where $\overline{Z}^{'} = \Sigma Z_{i} /n$ is the unweighted mean metallicity, 
and $n_{\rm eff}$, the effective number of DLAs, is given by 
\begin{equation}
n_{\rm eff} = {1 \over {2}} \biggl\{ 1 + \biggl[1 + {4 (n-1) [\Sigma (Z_{i} - 
\overline{Z}^{'})^{2} ] \over 
{\Sigma w_{i} (Z_{i} - \overline{Z})^{2}}} \biggr]^{1/2} \biggr\}. 
\end{equation}
If all the weights $w_{i}$ were equal, $n_{\rm eff}$ would be equal to $n$, 
$\overline{Z}^{'}$ would be equal to $\overline{Z}$, 
and $\sigma(\overline{Z})_{\rm samp}$  would be the usual 
standard deviation among the values $Z_{i}$, divided by $\sqrt{n}$. However, 
with unequal weights, the effective number of systems $n_{\rm eff}$ is 
smaller than $n$, and 
the weighted mean is dominated by only the few DLAs with the largest 
$N(\rm{H \, I})$. 

The other source of uncertainty in $\overline{Z}$ is   
the measurement error in the individual values of $N({\rm H \, I})$ and 
$N({\rm Zn \, II})$. The median measurement errors in the H I column densities 
are $20 \%$, but there is a wide range with $\approx 80\%$ of the values 
in the range $12-26 \%$. 
The median measurement errors in the Zn II column densities are $25\%$, 
with $\approx 80\%$ of the values in the range $7-40 \%$. (In general, 
the errors 
from observations obtained at eight-meter class telescopes are smaller than 
those from observations at four-meter class telescopes.) For any given DLA, 
the measurement errors in $N({\rm H \, I})$ and $N({\rm Zn \, II})$ are 
expected to be uncorrelated, since these quantities are derived from 
independent measurements of the H I Ly-$\alpha$ line and the Zn II lines.  
We therefore evaluate the effect of these individual measurement errors 
on the $N({\rm H \, I})$-weighted mean metallicity 
$\overline{Z}$, using   
the standard formula for error propagation:   
\begin{equation}
\sigma^{2}({\overline Z})_{\rm meas} = \Sigma \biggl[ {\partial {\overline Z} 
\over {\partial N({\rm Zn \, II})_{i} }} \biggr]^{2} 
\sigma^{2}[N({\rm Zn \, II})_{i}] + 
\Sigma \biggl[{\partial {\overline Z} \over {\partial N({\rm H \, I})_{i}}} 
\biggr]^{2} 
\sigma^{2}[N({\rm H \, I})_{i}]. 
\end{equation}
Using equations (3) and (13), we get
\begin{equation}
\sigma^{2}({\overline Z})_{\rm meas} = {[Z_{\odot} / {\rm (Zn/H)}_{\odot}]^{2} 
\Sigma \sigma^{2} [N({\rm Zn \, II})_{i}]  
+ {\overline Z}^{2} \Sigma \sigma^{2} [N({\rm H \, I})_{i}] 
\over {[\Sigma N({\rm H \, I})_{i}]^{2}}}.  
\end{equation}
The combined uncertainty in ${\overline Z}$ is given by 
\begin{equation}
\sigma^{2}({\overline Z})_{\rm tot} = \sigma^{2}({\overline Z})_{\rm samp} + 
\sigma^{2}({\overline Z})_{\rm meas}. 
\end{equation}
Finally, the uncertainty in the logarithmic $N({\rm H \, I})$-weighted mean 
metallicity is given approximately by 
\begin{equation}
\sigma[{\rm log} ({\overline Z}/Z_{\odot})]_{\rm tot} = 
{1 \over {2}} \biggl \{ {\rm log} 
\bigl[{\overline Z} + 
\sigma({\overline Z})_{\rm tot} \bigr] - {\rm log} 
\bigl[{\overline Z}-\sigma({\overline Z})_{\rm tot}\bigr] \biggr \}. 
\end{equation}

\subsubsection{The Maximum and Minimum Limits Cases}

To estimate the $N({\rm H \, I})$-weighted mean metallicity and its uncertainty 
in each redshift bin, we again consider 
two extreme cases allowed by the existing observations where the limits are 
treated either as detections (the ``maximum limits'' case) or as zeros 
(the ``minimum limits'' case).  These two cases provide strict upper and 
lower limits to the mean metallicity in each redshift bin, as 
verified by the survival analysis in section 2.2.2 below. 
Table 2 lists the results for each bin obtained in these two cases. 
Figure 1 shows the logarithmic $N({\rm H \, I})$-weighted mean metallicity 
relative to the solar value and its 1 $\sigma$ uncertainties, obtained 
as described above for our samples, as a function of redshift. Each point 
is plotted 
at the median redshift for the respective bin. Horizontal bars indicate 
the full range of redshifts of the DLAs in each bin. 
The left and right panels refer, respectively, to the 
cases of maximum and minimum limits. It is interesting to note 
that the mean metallicities and the relative trend  
with redshift do not change much whether the limits are treated as 
detections or zeros. This is because the mean metallicity in each bin is  
dominated by the high-$N({\rm H \, I})$ systems, for which the Zn values 
are invariably detections rather than limits. (This also clarifies that the 
intrinsic weakness of the Zn lines--compared to say Fe lines--
is not a serious problem in the estimation of the 
global metallicity.) The vertical error bars show the total uncertainties  
$ \sigma[{\rm log} ({\overline Z}/Z_{\odot})]_{\rm tot}$, including both 
the sampling errors, reflecting the 
scatter among individual DLAs, and the measurement errors in the 
column densities.  The sampling errors are found to 
be the dominant source of uncertainty, making up $\gtrsim 85 \%$ of the total 
error in each bin. 
The horizontal dashed line at ${\overline Z} = Z_{\odot}$ denotes the 
solar value. The curves in Figure 1 are the predictions of cosmic 
chemical evolution models and will be discussed in section 2.3. 
 
As for the unbinned approach, for easy comparison with the 
few previous studies that quantified the chemical evolution of DLAs, 
we fit a linear relation of the form of equation (8) to 
the observed logarithmic mean metallicity vs. redshift data. 
Based on linear regression, 
we obtain $ {\rm log} \,({\overline Z_{0} / Z_{\odot}})  = -0.76  
\pm 0.27$ and $b = -0.16 \pm 0.12$, counting the limits as detections, 
and $ {\rm log} \,({\overline Z_{0} / Z_{\odot}}) = -0.72 
\pm 0.30$ and $b = -0.23 \pm 0.15$, counting the limits as zeros. 
{\footnote{Unlike 
the case of the means in the individual redshift bins, the true ranges 
of the slope and the intercept are not necessarily bracketed by the minimum 
and maximum limits, but these two cases are indicative of the ranges 
the slope and the intercept are likely to take.}}  
These results are also summarized in Table 3. {\footnote{Recently, measurements 
of X-ray absorption lines have been used to estimate the metallicities of two 
DLAs at $z=0.313$ and $z=0.524$
(Bechtold et al. 2001; Junkkarinen et al. 2002). We have not included these in 
our sample, because there are no published Zn metallicities for these objects 
yet, and there could be a systematic difference between the Zn  
and X-ray-based metallicities. However, if we count these two systems in 
the lowest redshift bin, using the X-ray-based metallicities, the binned 
estimates of the intercept and the slope of the metallicity-redshift relation 
are ${\rm log} \,({\overline Z_{0} / Z_{\odot}})  = -0.30 \pm 0.21$ and 
$b = -0.36 \pm 0.10$ for the maximum limits sample.}}  $\chi^{2}_{\nu}$ 
for the optimum fits to 
the binned data is 0.47 counting the limits as detections, and 0.51 counting 
the limits as zeros, with the corresponding probabilities 
$P (\chi^{2}_{\nu} \ge \chi^{2}_{\nu, \, \rm{min}})$ of 0.93 and 0.92, 
respectively. 

\subsubsection{Survival Analysis}

Although the difference between the maximum limits and minimum limits cases 
is small, a more rigorous treatment of the limits can be achieved 
by using techniques of survival analysis for censored data (i.e. data 
with limits). 
We therefore also analyze the DLA Zn data using survival analysis. We use 
the Kaplan-Meier (K-M) non-parametric estimator, which is often 
used in biostatistics and has also been used in some astronomical applications 
with upper limits. We compute the mean metallicity in each redshift 
bin with the K-M estimator, using the astronomical survival analysis package 
ASURV rev. 1.2 (Isobe \& Feigelson 1990) which implements the methods 
presented in Feigelson \& Nelson (1985) for univariate problems. The last 
column in Table 2 lists the logarithmic $N({\rm H \, I})$-weighted mean 
metallicity for the five 
redshift bins, as obtained from survival analysis. As expected, these values 
lie in between the corresponding values for the maximum and minimum limits 
cases. 
{\footnote{The error bars on the mean metallicity in each bin 
for the survival analysis reflect only the sampling errors. We do not 
estimate the measurement errors in this case, because, to our knowledge, 
no method 
has been developed as yet for rigorously incorporating measurement errors 
into survival analysis with limits and detections. However, this is not a 
significant problem considering that  
$\gtrsim 85 \%$ of the contribution to the errors arises in 
the sampling errors for the maximum and minimum limits cases.}} 
Linear regression of these binned data yields an intercept 
$ {\rm log} \,({\overline Z_{0} / Z_{\odot}}) = -0.64 \pm 0.20$ and a slope  
$b = -0.26 \pm 0.10$ for the metallicity vs. redshift relation 
(see Table 3). 
The results of the survival analysis thus confirm the estimates from 
the cases of maximum and minimum limits. 

\subsection{Comparison of Data with Models}

The exponential form assumed above for the 
mean metallicity vs. redshift relation is of course an approximation, 
although convenient for purposes of comparison with previous studies. The 
true metallicity-redshift relation could be 
more complicated than this simple exponential form. 
Prochaska \& Wolfe (2000, 2002) and Prochaska et al. (2001a) have 
found a flat logarithmic mean metallicity vs. redshift relation for Fe for 
$2 \lesssim z \lesssim 4$. It may indeed be that the metallicity rises 
relatively slowly at high 
redshifts, but more steeply at low redshifts. With this in mind, we now 
compare the data with several models of 
cosmic chemical evolution. Very few of the previous studies have made 
even a graphical comparison between the models and the data. (The exceptions 
are Pei et al. 1999 and Savaglio 2001.)

The dotted, short-dashed, solid, and long-dashed curves in Figure 1 
show, respectively, (a) the mean interstellar metallicity in the closed-box 
and outflow models of Pei \& Fall (1995); 
(b) the mean Zn metallicity in the Malaney \& Chaboyer (1996) model with 
a constant Zn yield, a slope of $-1.70$ for the initial mass function 
(IMF), and the evolution of the neutral gas density 
as computed by Pei \& Fall (1995); (c) the mean interstellar 
metallicity in the Pei et al. (1999) model with the optimum fit for the 
cosmic infrared background intensity; and 
(d) the mean metallicity of cold interstellar gas in the 
collisional star-burst model of Somerville, Primack, \& Faber (2001). 
The metallicity evolution of 
the models clearly depends on several input parameters, such as the IMF slope, 
the Zn yields, and the star formation history of the galaxies. 
All of these models have near-solar metallicity at the present epoch, as 
required by the observed present-day mean interstellar metallicity of 
galaxies (see the Appendix). 
However, the models with shallower slopes tend to lie above the DLA data 
at $z \sim 2$, while the models with steeper slopes tend to go through the 
DLA data at $z \sim 2$. 

In the redshift range where the data and the models can be compared, 
the average slopes for the Pei \& Fall (1995)  models are $-0.54$ to $-0.42$, 
while the slopes for the Malaney \& Chaboyer 
(1996) models are 
$-0.45$ to $-0.25$. The models of Pei et al. (1999) 
predict slopes of $-0.61$ to $-0.45$. The hydrodynamical simulations of 
star formation and metal production in hierarchical clustering scenarios 
by Tissera et al. (2001) predict a slope of $-0.33$ for the mass-weighted 
mean logarithmic metallicity as a function of redshift. The semi-analytic 
collisional star-burst model for galaxy formation in the 
cold dark matter cosmogony by Somerville et al. (2001) predicts a 
slope of $-0.21$ for the mean metallicity of cold interstellar gas 
as a function of redshift (but predicts higher 
metallicities at all redshifts than observed for DLAs). The slopes of 
the observed mean metallicity-redshift relation (from Table 3) are marginally 
consistent with no evolution, but also agree, within 
$\approx 2 \sigma$, with most of the model predictions. 

To quantify the comparison of the data with the models of cosmic chemical 
evolution, we now compare the weighted $\chi^{2}$ for the unbinned data, 
following equation (6), with $\overline Z_{\rm p}$ in this case given by 
the cosmic chemical evolution models. 
The weights $w_{i}$ are, once again, the fractional contributions of the 
individual DLAs to the total $N ({\rm H \, I})$ of the entire sample. 
As before, we calculate $\sigma_{Z}$ as the scatter 
in the data with respect to the best exponential fit  
(as determined in section 2.1).  
The $\chi^{2}$ thus defined cannot give us independent estimates of 
the goodness of fit of each model. But the relative values of $\chi^{2}$ 
can be used to compare how well the models fit the data. 

Table 4 lists $\chi^{2}_{\nu}$ for each of the four 
models shown in Figure 1 with respect to the unbinned data. Also listed 
are the corresponding values for the  Pei et al. (1999) models  
for the maximum and minimum allowed fits to the cosmic 
infrared background intensity. On the same scale, the corresponding value 
of $\chi^{2}_{\nu}$ for the best exponential fit is 1 by definition. 
Even though the $\chi^{2}_{\nu}$ values in Table 4 do not reflect the 
goodness of fit in an absolute sense, they are roughly similar 
to the values expected on the basis of the comparison of the curves in 
Figure 1 with the binned data points. The main reason for the higher values of 
$\chi^{2}_{\nu}$ for the chemical evolution models than for 
the exponential fit is that 
the models lie above the data points. The Malaney \& Chaboyer (1996) 
and Somerville et al. (2001) models 
have shallower slopes than the other models, but their offset to 
higher metallicities give them worse $\chi^{2}_{\nu}$, as 
evident from Figure 1. Overall, we conclude that at least some of the 
chemical evolution models are consistent with the DLA data, supporting    
some evolution in the global metallicity as a function of redshift. 
We note, however, that dust in DLA galaxies can introduce a systematic 
selection bias in the observed mean metallicity. The curves plotted in 
Figure 1 have not been corrected for this bias. We discuss this point 
further in section 3.1. In future, when more measurements at low 
redshifts become available, it should be possible to make more 
stringent comparisons of the models with the data. 

\section{DISCUSSION}

Most previous studies have claimed that there is no 
evolution in the global metallicity of DLAs. However, these studies 
have not been definitive for various reasons summarized in Table 1. 
Our analysis has demonstrated that the present DLA Zn data are 
consistent with some evolution of the global interstellar metallicity.  
The main reason for the large uncertainties is that 
the effective number of measurements that dominate the 
$N({\rm H \, I})$-weighted  mean metallicity is very small. 

A complete lack of evolution in the mean metallicity would 
be quite surprising. Such a trend  
would be hard to reconcile with the inference that the global rate of 
star formation was high at $1 \lesssim z \lesssim 4$, based on the luminosity 
density of galaxies observed in deep surveys such as the Canada-France 
Redshift Survey and the Hubble Deep Field (e.g. Lilly et al. 1996; Madau 
et al. 1996, 
1998). The metallicity of interstellar matter in galaxies is thus expected 
to 
rise with time. Indeed, the census of the metals in nearby 
galaxies shows a luminosity-weighted mean interstellar metallicity close to 
the solar value. (See the Appendix.) 
DLAs are believed to represent the interstellar matter of galaxies, and are 
therefore expected to show an increase in the mean metallicity 
with decreasing redshift. As we have shown, the DLA data are, in fact,  
consistent with the metallicity evolution predicted by some cosmic 
chemical evolution models. 

\subsection{Dust Obscuration Bias}

Dust in DLA galaxies could bias the empirical estimates of 
the mean metallicity $\bar Z$ and its evolution with redshift.
This is because the DLA galaxies with the highest column densities of 
metals are likely to be those with the highest column densities of dust,
and these may obscure background quasars to such a degree that
some of them are omitted from optically selected samples (Fall 
\& Pei 1993; Boisse et al. 1998).
Evidence for dust in DLA galaxies comes from the statistical
reddening of background quasars (Fall, Pei, \& McMahon 1989;
Pei, Fall, \& Bechtold 1991) and the depletion patterns of
heavy elements, especially Zn and Cr (e.g., Pettini et al. 1994, 
1997). 
Like the mean metallicity, the mean dust-to-gas ratio in the 
observed DLA galaxies is low at high redshifts, where most of
the measurements have been made.
The mean dust-to-metals ratio appears to be roughly independent 
of redshift and about equal to that in the Milky Way and the 
Magellanic Clouds (see Figure 1 of Pei et al. 1999).
As a result of obscuration, the observed mean metallicity in 
the DLA galaxies, and hence the data points in Figure 1, may 
lie systematically below the true mean metallicity.
The curves in Figure 1, however, are predictions for the true 
mean metallicity, without corrections for obscuration, from 
the models of cosmic chemical evolution.

The severity of this bias depends on several factors, including 
the extinction curve of the dust, the distribution of dust 
column densities in the DLA galaxies, the 
luminosity function of quasars, and the passband in which 
they are observed (see the Appendix of Fall \& Pei 1993 for 
a detailed analysis). Together, these factors determine the fraction  
of the sky covered by dust as a function of the optical depth. 
The main difficulty in correcting for the bias stems from the 
unknown distribution of dust column densities in the DLA galaxies or, 
for a given (observed) distribution of H I column densities, the unknown 
distribution of dust-to-gas ratios.
For an assumed shape of this distribution, one can, however, 
compute the expected bias as a function of the width of the 
distribution.
Fall \& Pei (1993) have made such calculations for a log-normal
distribution of the dust-to-gas ratio $k$ in the DLA galaxies.
They find that the true mean dust-to-gas ratio $\bar k_t$ exceeds 
the observed mean $\bar k_o$ by a factor that increases from 
$\bar k_t/\bar k_o = 1.2$ to 1.7 to 4 as the dispersion in the 
natural logarithm of the dust-to-gas ratio increases from 
$\sigma(\ln k) = 0.5$ to 1.0 to 1.5.
This may provide an indication of the corresponding bias in
the mean metallicity, since we expect $\bar Z_t/\bar Z_o 
\approx \bar k_t/\bar k_o$.
For reference, nearby normal galaxies have $\sigma (\ln k) \approx 
0.5$ (Pei 1992).
Thus, if this dispersion also applies to the DLA galaxies, we 
might expect $\bar Z_t \approx \bar Z_o$.
If the dispersion in the dust-to-gas ratio were larger, however, 
$\bar Z_t$ could differ significantly from $\bar Z_o$.

Unfortunately, the bias caused by obscuration is difficult 
to quantify from theory or numerical simulation alone,  
because it depends on the the small-scale 
structure of the interstellar medium in the DLAs. 
Absorption-line observations sample the DLAs on scales 
comparable to the continuum-emitting regions of the 
background quasars, typically smaller, and possibly much 
smaller, than a light-year. Some of these lines of sight 
will pass through dense interstellar clouds, with high 
optical depths, while others, even nearby, will pass 
through diffuse intercloud material, with low optical 
depths. The obscuration of background quasars in this
case may differ substantially from that in simulations 
with the same average interstellar density and hence 
optical depth but with lower spatial resolution and
hence little or no structure on the relevant scales. 

In principle, a comparison of metal abundances in DLAs from optical vs. 
radio-selected quasars can help to quantify the dust selection effect. 
Ellison et al. (2001) find slightly more DLAs in the foreground of 
radio-selected and optically faint quasars than in optically 
selected and optically bright quasars, in the sense that may be 
caused by dust obscuration. However, the sample of radio-selected 
quasars is still small, and neither of these differences is statistically 
significant. The strongest conclusion 
that can be drawn at present is that obscuration reduces  
the number density and/or $\Omega_{\rm H \, I}$ of DLAs at 
$1.8 < z < 3.5$ in optically selected samples by a factor of two or less. 
However, the implications 
of this result for the mean metallicity 
have not yet been quantified. Abundance studies for a large sample of DLAs in 
radio-selected and optically faint quasars will help to improve the 
constraints on the extent of the dust obscuration bias. 

\subsection{Iron vs. Zinc}

All of our analysis has been based on Zn alone, since  
we believe Zn is a more reliable metallicity indicator for DLAs than Fe, 
which has been used in some other studies (e.g., Prochaska \& Wolfe 1999, 
2000; Savaglio 2001; Prochaska et al. 2001a). The main reasons advocated 
by these studies for using Fe are: (a) the ease of measuring Fe lines 
compared to the weaker Zn lines; (b) the ability to probe 
higher redshifts ($z > 3.5$) with Fe than with Zn; and (c) the better 
understanding of the nucleosynthetic origin of Fe than of Zn.  
We address these issues one by one. (a) 
As our ability to measure weak absorption lines is getting better 
in the present age of large telescopes, Zn measurements are 
becoming easier. Furthermore, this is not a concern for studies of 
the global $N({\rm H \, I})$-weighted metallicity as the systems that are 
too weak to give detections of Zn lines do not contribute much to the 
global metallicity anyway. As our analysis has shown, 
there would be almost no difference in the results even if the 
limits could be improved in future studies (as there is little 
difference whether the limits are treated as detections or zeros). 
(b) The redshift range $z > 3.5$ accessible with Fe is of some 
interest since the rate of metallicity evolution at high redshifts 
could in fact be 
different from that at low redshifts. However, the redshift range $z > 3.5$ 
represents only $13 \, \%$ of the cosmic 
history. Therefore, even though Fe is useful for tracing the early  
chemical evolution of galaxies, it does not add much to studies of 
the last $87 \, \%$ of the age of the universe. 
(c) While the nucleosynthetic origin of Zn is harder to understand 
than that of Fe, observationally Zn and Fe track each other perfectly 
well for most Galactic disk and halo stars with metallicities between 
$10^{-2}$ solar and solar, the metallicity range relevant to DLAs 
(e.g., Sneden, Gratton, \& Crocker 1991). 
Some models of massive star explosive nucleosynthesis, such as 
the neutrino-driven wind models (e.g., Hoffman, Woosley, \& Qian 1997) provide 
ways to understand the origin of Zn and the reason for the tight 
correlation between Fe and Zn in Galactic stars. Thus, there is no strong 
reason to suspect that the use of Zn biases studies of cosmic chemical 
evolution in any significant way. 

On the other hand, the well-known problem with Fe is its strong 
depletion on dust grains. Moreover, as with many other elements, the 
depletion of Fe differs substantially between the cool and warm diffuse 
interstellar gas, but is strong in both phases 
(see, e.g., Savage \& Sembach 1996). By comparison, Zn is essentially  
undepleted in the warm interstellar clouds and depleted $\sim 40$ times 
less than Fe in the cool interstellar clouds. It is difficult to 
model unambiguously the dust depletion 
effects for Fe, because the structure and composition of the dust 
grains present in the DLAs is not known a priori. Furthermore, 
the line of sight can pass through a mixture of warm and cold gas, 
so that the dust depletion within a given DLA can be quite different 
in different 
parts of the absorbing gas. Even after averaging over a number of 
DLAs at a given redshift, it is possible that the mean correction for  
dust depletion may itself change as a function of redshift. 
Indeed, as the interstellar metallicity of the DLA galaxies 
rises with decreasing redshift, their dust-to-gas ratio should also 
rise more or less in step with the metallicity. Such a 
redshift-dependent dust depletion can introduce an error in estimates of 
the metallicity-redshift relation 
based on Fe data alone, or on a combination of Fe and Zn data. Although 
the extent of this error is hard to quantify and, in fact, may turn out 
to be small, it is safer to focus on an element such as Zn that 
does not have this problem in the first place. 

\subsection{Future Work}

Part of the reason our analysis finds a stronger evidence for evolution 
than previous studies such as Pettini et al. (1999) is the addition of 
new data at low redshifts that have recently become available. These new 
data have resulted  
in a higher mean metallicity in the lowest redshift bin than found 
in the earlier analyses. This highlights the need for caution in drawing 
conclusions from the limited DLA data sets. Indeed, it is 
obvious that the current samples need significant improvement 
in the number of measurements at $z < 2$. A large fraction of 
the Zn measurements so far have focussed on the redshift range $z > 2$. 
Possible drops in the 
mean Zn metallicity at $3 < z < 3.5$ and the mean Fe metallicity 
at  $3.5 < z < 4.5$ have been suggested (although on the basis of 
small samples) by Pettini et al. (1997) and Prochaska \& Wolfe (2002), 
respectively. 
If verified with future data, a drop in the mean metallicity at 
high redshifts could signal the epoch of the onset of star formation in 
DLAs. It is, nevertheless,  
also essential to increase substantially the lower redshift samples, 
since the redshift range $z < 2$ probes the cosmic epochs when the bulk 
of the metals in galaxies were produced. At present, Zn measurements 
exist for only seven absorbers at 
$z < 1$ and only two absorbers at $z < 0.5$. This is especially problematic 
because the redshift ranges 
$z < 1$ and $z < 0.5$ represent $57 \, \%$ and $37 \, \%$, respectively,  
of the age of the universe. Even the number of measurements at 
$1 < z < 2$ is somewhat limited. Clearly, it is crucial to 
increase the number of Zn measurements at low and intermediate 
redshifts. This requires more measurements 
at $z < 0.6$ with the {\it Hubble Space Telescope} (HST) and at 
$0.6 < z < 2$ with ground-based telescopes. 
It is particularly important to study the high-$N({\rm H \, I})$ systems 
since, as we have emphasized here, these systems dominate the global 
metallicity. 

The large numbers of quasar spectra becoming available with surveys such as 
the Sloan Digital Sky Survey and the FIRST survey will 
increase the sample of known DLAs and 21-cm absorbers by a large factor. 
Obtaining element abundances of these new DLAs will be of great importance 
for pinning down the metallicity-redshift relation for DLAs. 
Zn abundances in  DLAs in front of quasars of different apparent 
magnitudes and colors will provide constraints 
on the amount of dust obscuration. Accurate determination of the 
metallicity-redshift relation for 
DLAs will, thus, be important for quantifying the selection effects in 
samples of quasar absorbers and for understanding the nature of DLAs, 
in addition to providing important 
constraints on the global star formation history of galaxies. 

\acknowledgments
We thank Eric Feigelson for providing the astronomical survival analysis 
package ASURV and for helpful discussions. We thank Hsiao Wen Chen, 
Edward Jenkins, Max Pettini, Jason X. Prochaska, and an anonymous 
referee for helpful comments. 
VPK acknowledges partial support from a Research and Productive Scholarship 
award from the University of South Carolina, from the University of South 
Carolina Research Foundation, and from NASA/South Carolina 
Space Grant Consortium. SMF thanks the Carnegie Observatories for 
hospitality during the later phases of this project. 

\appendix
{\section{APPENDIX}
Here, we estimate the global mean interstellar metallicity of galaxies 
at the present epoch from emission-line observations of H II regions. 
As an approximation to the mass-weighted mean metallicity, required 
for models of cosmic chemical evolution, we compute the luminosity-weighted 
mean metallicity, since the latter is much simpler to evaluate than the 
former. 
The luminosity-weighted mean interstellar metallicity of galaxies can be 
expressed in the form 
\begin{equation}
{\overline Z} = 
{{\int Z(L)\, L \, \phi (L) \, dL } \over {\int L \, \phi(L) \, dL}},
\end{equation}
where $Z(L)$ is the average interstellar metallicity in a galaxy of luminosity
$L$, and $\phi (L)$ is the luminosity function of galaxies of all 
morphological 
types. To evaluate equation (A1), we adopt a 
Schechter function for $\phi (L)$ and a power law for $Z(L)$, i.e.,  
\begin{equation}  
\phi(L) dL = \phi^{*} (L/L^{*})^{\alpha} {\rm exp}(-L/L^{*}) dL / L^{*}, 
\end{equation}
and
\begin{equation}
Z(L) = (L/L^{*})^{\beta} Z^{*}, 
\end{equation}
where $Z^{*}$ is the mean 
metallicity of galaxies with luminosity $L^{*}$. Thus, we obtain 
\begin{equation}
\overline Z = {\Gamma(\alpha + \beta + 2)
\over {\Gamma(\alpha + 2)}} Z^{*}. 
\end{equation}
For the parameters of the luminosity function, we adopt the results of 
Loveday et al. (1992), i.e., $\phi^{*} = 5.9 \times 10^{-3} \, h_{75}^{3}$ 
Mpc$^{-3}$, 
$\alpha = -0.97$, and $L^{*} = 2.3 \times 10^{10} \, h_{75}^{-2} \, L_{\odot}$ 
(corresponding to 
$M^{*}_{B} = -20.42$ for $H_{0} = 75 \, h_{75}$ km s$^{-1}$ Mpc$^{-1}$). 
For the relation between mean metallicity and luminosity, we adopt the 
parameters $Z^{*} = 0.90 \, Z_{\odot}$ and $\beta = 0.42$, which provide an 
adequate fit to the data for [O/H] from H II regions vs. $M_{B}$ in a sample 
of 39 nearby spiral galaxies from Zaritsky, Kennicutt, 
\& Huchra (1994) and 20 irregular galaxies from Skillman, Kennicutt, \& Hodge 
(1989) and references therein. 
Inserting these parameters in equation (A4), we obtain the mean 
interstellar metallicity of nearby galaxies $\overline Z  \approx 0.8 \, Z_{\odot}$. 

\clearpage

%
%

\clearpage
%
%
\begin{figure}
\plottwo{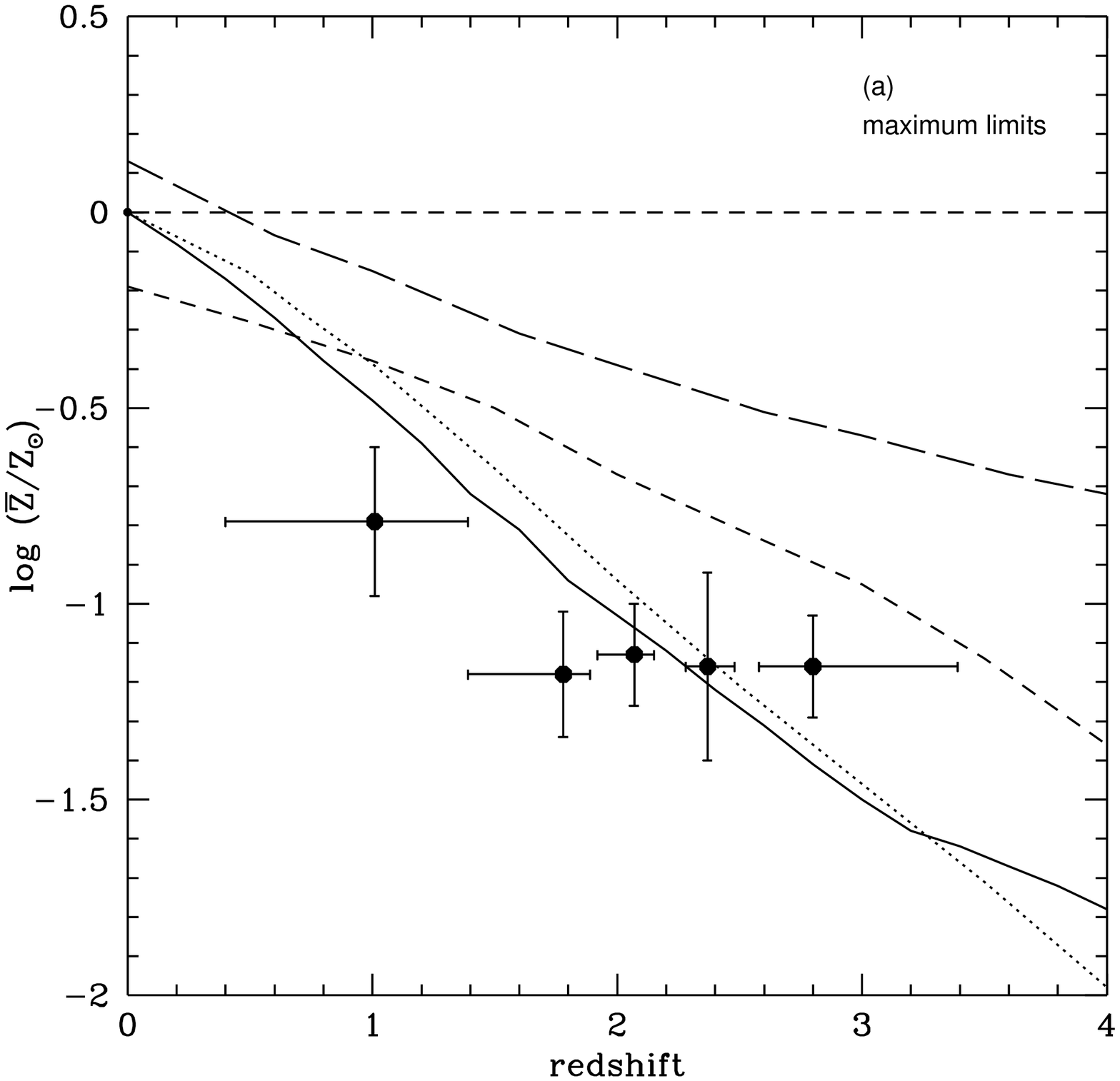}{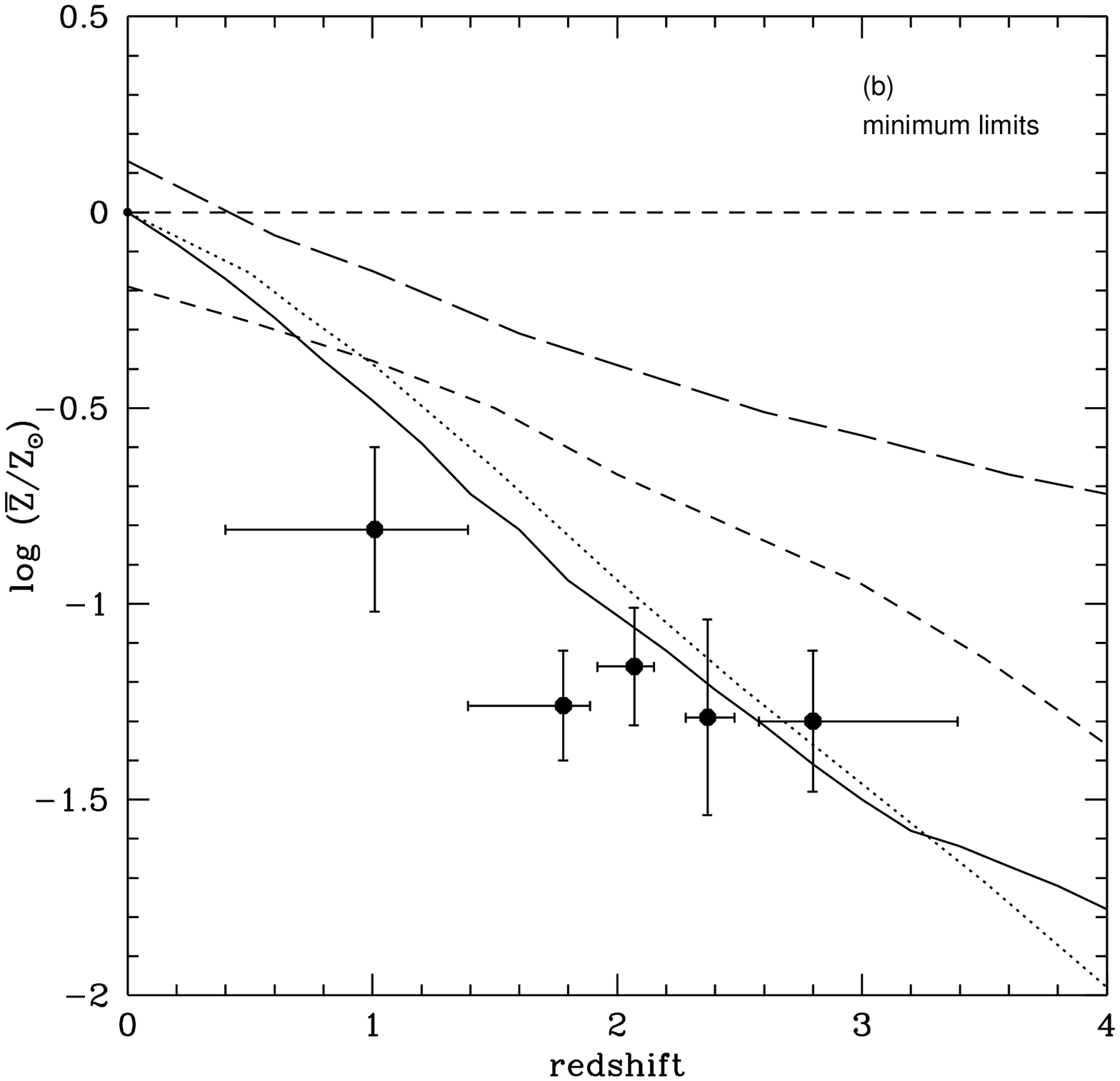}
\caption{The global metallicity-redshift relation deduced from 
damped Ly-$\alpha$ absorbers. Filled circles show the logarithmic 
$N({\rm H \, I})$-weighted mean Zn metallicity relative to the solar value  
vs. redshift for measurements from the literature. 
Left panel (a) includes detections and upper limits treated as detections, 
while right panel (b) includes detections and upper limits 
treated as zeros. Vertical error bars denote 1 $\sigma$ uncertainties 
in the logarithmic $N({\rm H \, I})$-weighted mean metallicity. 
Data points are plotted at the median redshift in each bin. 
Horizontal bars denote the full range of 
redshifts of the DLAs in each bin. Horizontal dashed line at 
${\overline Z} = Z_{\odot}$ denotes the solar level.
Dotted, short-dashed, solid, and long-dashed 
curves show, respectively, the `true' mean metallicity (not corrected for dust 
obscuration) expected in the cosmic chemical evolution models 
of Pei \& Fall (1995),  Malaney \& Chaboyer (1996), Pei et al. (1999), 
and Somerville et al. (2001). See text for further details. \label{fig1}}
\end{figure} 

\clearpage

\begin{table}
\centerline{\bf {TABLE 1}}
\centerline{\bf{Previous Studies of Global Metallicity Evolution in DLAs}}
\vspace{1ex}
\begin{center}
\begin {tabular}{llccccccc}
\tableline
\tableline
Ref.$^{\dagger}$ & Element&Detections, &$z$ range & Data&Errors in&
Slope& Evolution?\\
&&Limits,&&binned?&weighted &fitted?&\\
&&Treatment$^{\dagger \dagger}$&&& $\overline{Z}$?&&&\\
\tableline
1 & Zn & 19, 15, D&$0.7-3.4$&Yes & No & No & No  \\
2  & Zn & 24, 16, D&$0.4 -3.4$& Yes & No & No & No  \\
3 & Zn & 11, 4, D&$ 1.8-2.6$ &Yes&No &No& No  \\
3 & Fe & 19, 1, D& $1.8 -4.2$ &Yes &No &No& No \\
4 & Zn & 28, 0, E& $0.4 - 3.4$ & Yes & No & $-0.13$& No \\
&&&&&&$\pm 0.07$&\\
5 & Fe & 39, 0, E &$1.8 -4.2$&  Yes & Yes & No & No \\
6 & Zn, Fe$^{*}$ & 75, 0, E&$0 -4.4$&  Yes & No & No& No  \\
7 & Fe$^{*}$ & 35, 0, E &$ 1.8- 4.5$ &  Yes & Yes & No& No \\
8 &Fe & 51, 0, E &$1.6 -4.5$ &  Yes & Yes & No& No \\
8 &Zn & 10, 5, D &$1.6 -2.8$ &  Yes & Yes & No& Possible$^{**}$\\
\tableline
This &Zn &36, 21,   & $0.4 -3.4$& Yes, & Yes & $-0.26^{***}$ & 
Likely\\
work&&D,Z,S&&No&&$\pm 0.10$&\\
\tableline
\end{tabular}
\end{center}
\noindent{ $\dagger$ References: 1. Pettini et al. (1997); 2. Pettini et al. 
(1999); 
3. Prochaska \& Wolfe (1999); 4. Vladilo et al. (2000); 5. Prochaska \& Wolfe 
(2000); 
6. Savaglio (2001); 7. Prochaska, Gawiser, \& Wolfe (2001a); 8. Prochaska \& 
Wolfe (2002).}

$\dagger\dagger$ D: Limits treated as detections; E: Limits Excluded; 
Z: Limits treated as zeros; S: Limits treated with survival analysis. 

*: Sometimes other elements (Cr, Ni, Si etc.) used as a proxy for Fe or Zn. 

**: But Prochaska \& Wolfe (2002) state that this may be because of 
the small sample size.

***: Value obtained with survival analysis.

\end{table}
\clearpage
\begin{table}
\centerline{\bf {TABLE 2}}
\centerline{\bf{Global Zn Metallicity vs. Redshift (Binned)}}
\vspace{1ex}
\begin{center}
\begin {tabular}{cccccc}
\tableline
\tableline
$z$ range & Detections,& Median $z$ 
&${\rm log}\,(\overline Z / Z_{\odot})$&${\rm log}\,(\overline Z / Z_{\odot})$&
${\rm log}\,(\overline Z / Z_{\odot})$\\
&Limits& & Max. Limits & Min. Limits&Survival Analysis\\
\tableline
0.40-1.39&9, 2 &1.01 &$-0.79 \pm 0.19$&$-0.81 \pm 0.21$& $-0.80 \pm 0.12$\\
1.39-1.89&7, 4&1.78&$-1.18 \pm 0.16$& $-1.26 \pm 0.14$& $-1.23 \pm 0.10$\\
1.92-2.15&8, 4&2.07&$-1.13 \pm 0.13$& $-1.16 \pm 0.15$&$-1.16 \pm 0.15$\\
2.28-2.48&8, 3&2.38&$-1.16 \pm 0.24$&$-1.29 \pm 0.25$&$-1.22 \pm 0.09$\\
2.58-3.39&6, 6&2.80&$-1.16 \pm 0.13$&$-1.30 \pm 0.18$&$-1.25 \pm 0.20$\\
\tableline
\end{tabular}
\end{center}
\end{table}
\clearpage
\begin{table}
\centerline{\bf {TABLE 3}}
\centerline{\bf{Exponential Fits to Global Metallicity-Redshift Relation}}
\vspace{1ex}
\begin{center}
\begin {tabular}{cccc}
\tableline
\tableline
Binning&Treatment of Limits &${\rm log}\,(\overline Z_{0} / Z_{\odot})
$&$b$ \\
\tableline
Unbinned &Maximum limits &  $-0.71 \pm 0.20$ & $-0.20 \pm 0.11$\\
Unbinned &Minimum limits &   $-0.64 \pm 0.20$ & $-0.27 \pm 0.12$ \\
Binned & Maximum limits &  $-0.76 \pm 0.27$ & $-0.16 \pm 0.12$\\
Binned & Minimum limits &  $-0.72 \pm 0.30$ & $-0.23 \pm 0.15$ \\
Binned & K-M Survival Analysis &  $-0.64 \pm 0.20$ & $-0.26 \pm 0.10$ \\ 
\tableline
\end{tabular}
\end{center}
\end{table}
\clearpage
\begin{table}
\centerline{\bf {TABLE 4}}
\centerline{\bf{Fits of Chemical Evolution Models to 
Metallicity-Redshift Relation}}
\vspace{1ex}
\begin{center}
\begin {tabular}{lcc}
\tableline
\tableline
Model &Limits& $\chi^{2}_{\nu}$ \\
\tableline
Pei \& Fall (1995) &Maximum Limits &2.69 \\
&Minimum Limits &3.07 \\
Malaney \& Chaboyer (1996) &Maximum Limits & 3.67 \\
&Minimum Limits &  4.47 \\
Pei et al. (1999) Maximum IRB & Maximum Limits &  2.13 \\
&Minimum Limits & 2.40 \\
Pei et al. (1999) Optimum IRB & Maximum Limits &  1.97 \\
&Minimum Limits & 2.18 \\
Pei et al. (1999) Minimum IRB & Maximum Limits &  1.76 \\
&Minimum Limits & 1.87 \\
Somerville et al. (2001) & Maximum Limits & 14.54 \\
&Minimum Limits & 17.65 \\
\tableline
\end{tabular}
\end{center}
\end{table}
\end{document}